\def\beg{\begin{equation}}
\def\eeq{\end{equation}}
\documentstyle[12pt]{article}
\begin{document}
\begin{center}
{\Large{\bf The composite Fermion model of quantum Hall effect is internally inconsistent}}
\vskip0.35cm
{\bf Keshav N. Shrivastava}
\vskip0.25cm
{\it School of Physics, University of Hyderabad,\\
Hyderabad  500046, India}
\end{center}

The composite fermion (CF) model of the quantum Hall effect which gives the correct series of charges is based on attachment of flux quanta to the electron. The construction of the series of charges leads to a field expression which requires that flux quanta are attached to the electron. The series based on the experimental data is correct but the field deduced from such a series is found to be incorrect. The size of the CF is compared with the electron radius and it is found that for the same density, the CF are internally inconsistent. The attachment of the flux quanta to the electron or detachment of flux quanta from CF is neither found experimentally nor
is feasible theoretically.
\vfill
Corresponding author: keshav@mailaps.org\\
Fax: +91-40-3010145.Phone: 3010811.
\newpage
\baselineskip22pt
\noindent {\bf 1.~ Introduction}

     The fractional charges found in the experimental data of quantum Hall effect are described by $e_{eff} = \nu e$ where two expressions for $\nu$ appear to give the correct experimental values,
\beg
\nu_+ = {\nu^*\over 2p\nu^*+1}
\eeq
and
\beg
\nu_- = {-\nu^*\over 2p\nu^*-1}
\eeq
The above expressions are symmetric with respect to interchange of
$\nu$ and $\nu^*$ so that,
\beg
\nu^* = {\nu \over 2p\nu+1}, \, \, \, \nu^*={-\nu\over 2p\nu-1}.
\eeq
For p=1, the first of these series gives,
\beg
\nu^*={\nu\over 2\nu+1}
\eeq
which is the same as found experimentally for $e_{eff}=\nu^*e$ and integer $\nu$. For $\nu$ =1, 2, 3, 4, etc. correct values of the effective charge are found.

     Jain(1989, 1998) introduced two types of quasiparticles, the composite fermions (CF) and the electrons. The magnetic field is quantized for both of these quasiparticles but they see different fields,
\beg
\nu ={\rho\phi_o\over B} (electrons);
\eeq
and
\beg
\nu^*={\rho\phi_o\over B^*}(CFs).
\eeq
The two fields are related by the expression,
\beg
B^*= B - 2p\rho\phi_o.
\eeq
If we put B$^*$=0, then B is quantized and when we put B=0, then B$^*$ is quantized. It is assumed that $\phi_o$ can be parallel as well as antiparallel to B. Here p is an integer, $\nu$ and $\nu^*$ are also integers. By substituting integer $\nu$, fractional $\nu^*$ is produced. The number density of electrons per unit area is $\rho$ and $\phi_o$=hc/e. While $\nu^*=\nu/(2\nu+1)$ gives a series of experimentally measured values correctly, the field $B^*$ is not observed experimentally. Therefore, we wish to see if (7) is internally consistent. The factor 2p is an even number and it produces the correct series (4), i.e., the factor 2 in the denominator of (4) is correct. According to the expression (7), even number of flux quanta are attached to the electron, i.e., 
\beg
CF = e + 2p \,\,vortices.
\eeq
We can represent this quasiparticle by a picture showing an electron with two (even number) of flux quanta attached as shown in Fig.1.

\noindent{\bf 2.~~Comments}

(a). We find that the series (4) is experimentally observed but the field (7) is not observed experimentally. If there was a phase transition at which the even number of flux quanta were attached while cooling and detached while heating, then it was satisfactory to represent the field by (7) but no such phase transition is found experimentally.

The number density of electrons is given by $\rho$ = n$_o/{\it l}^2$, where {\it l} is the magnetic length, so that for $\nu$=1,
\beg
{\it l} = {(n_o\phi_o/ B)}^{1/2}.
\eeq
The size of the electron may be determined by the classical radius of the electron which is given by, $e^2/mc^2$,
\beg
e^2/mc^2 = 2.8179 \times 10^{-13} \,\,cm
\eeq
or by the Compton wave length which is the inverse mass,
\beg
\hbar/mc = 3.86159 \times 10^{-11}\,\,cm
\eeq
For h=6.6262$\times$ 10$^{-27}$ erg$\,\,$s, c= 2.9979 $\times$ 10$^{10}$ cm/s and e = 4.80325$\times$ 10$^{-10}$ esu, $\phi_o$=hc/e=4.13 $\times$ 10$^{-7}$ G$\,\,$cm$^2$ so that at B=28.6 T, n$_o$=1,
\beg
{\it l}= 1.2 \times 10^{-4} cm.
\eeq
Therefore, the magnetic length is seven orders of magnitude 
larger than the Compton wave length of the electron. Similarly, 
the magnetic length is nine orders of magnitude larger than the classical radius of the electron. Therefore, if flux quanta are 
attached to the electron, the CF are going to be vary large 
objects while the electrons are small. In the expressions  (5)
 and (6) the density of CF is equal to the density of electrons. 
Why the very large CFs should have the same  density as that of electrons? A small number of large objects can fill the same 
space as a large number of small objects but
 large objects can not fill the same space with the same number
 as the small objects
unless the larger ones are squeezed, 
 but there is no provision to squeeze in any of the formulas.
 Therefore, the expressions (5) and (6) are internally
 inconsistent with (7). There is not enough room to attach a 
lot of vortices to electrons. When a large number of vortices
 are attached to the electrons, we should require a lot more
 space than that occupied by electrons. Therefore CF model is
 internally inconsistent.

\noindent(b) When electrons get attached to the flux quanta, 
let us say 2, then the resulting CF is a large object so that
 they push some electrons out of the sample. Thus the charge 
leaks out of the sample resulting into flow of electrons or a 
current. No such current has been detected in the experiments.
 However, the leak current will disturb the resistivity
 measurement. If a constant voltage has to be maintained,
 the reduced current has to be balanced by an increase in
 resistivity in addition to that given by the classical Hall
 effect. Over and above the Hall resistivity, an additional 
increase in the resistivity should occur but no such increase
 in resistivity has been detected. Alternatively, the current
 drop should reflect in the voltage drop but no such voltage 
drop has been detected experimentally. Therefore, it can be said
 safely that composite fermions (CFs) have not been seen
 experimentally and the claims made are incorrect. According
 to the CF model, flux quanta are attached to the electron but
 it is found that no such flux quanta are attached to the 
electron. The attachment of flux quanta to electrons has 
not been found in the experimental data.

     It has been pointed out by Kumada et al that there are
 two types of quasiparticles in the flanks of $\nu$=2/3. 
Hysteresis can arise because the energies of these 
quasiparticles having opposite spins depend on Zeeman energy.
 In fact "the opposite spin" aspect is not a part of the CF
 model and this feature observed by Kumada et al is not 
contained in the CF model. In fact this experimental feature
 is obtained only in Shrivastava's theory. Kumada et al make
 a case for CF on the basis of Fermi liquid, activation gap, polarization, cyclotron resoance, etc. but none of these 
experiments prove that {\it flux quanta are attached} to 
the electrons. The Fermi liquid does not require the flux
 attachment. The activation is an old barrier problem not
 connected to the flux attachment. The polarization where it
 has been measured correctly by NMR does not require flux
 attachment. Similarly, the cyclotron resonance can be done
 without attaching flux quanta to the electrons. The CF 
theory also violates the Biot-Savarts law of classical 
electrodynamics and hence CF is not consistent with Maxwell 
equations. The CF also do not obey the flux quantization
 correctly.
\vskip0.25cm
\noindent{\bf 3.~~Proper theory}

     The correct theory of the quantum Hall effect is given by Shrivastava$^4$. It has been pointed out that all of the 
experimental data on the quantum Hall effect agrees with 
the theory of Shrivastava.$^4$
\vskip0.25cm

      We are not ignoring the award of Oliver Buckley Prize 
to Jain$^1$
for the CF model but our paper is dated earlier. Some 
comments have appeared in the literature$^6$. Willett's$^7$
 data is also fully in agreement with the theory of
 Shrivastava$^4$. Similarly, the "opposite spin" feature
 pointed out by Kumada et al is clearly demonstrated in 
ref.4, but not in CF model.

\noindent{\bf4.~~Conclusions.}

We find that the flux quanta are not attached to the electrons. 
Hence the CF model is incorrect. The flux attachment is neither 
found experimentally nor is feasible theoretically. Such experimentalists who are claiming to have observed the CF have
 observed only the series of charges but not the flux attachment. Usually it is a custom to agree with the decisions and proceed 
by assuming that the ``awarded result" is correct but in the case
 of CF the award is clearly misplaced.\\

\noindent{\bf5.~~References}
\begin{enumerate}
\item J.K. Jain, Phys. Rev. Lett. {\bf63}, 199 (1989).
\item K. Park and J.K. Jain, Phys. Rev. Lett. {\bf81}, 4200 (1998).
\item N. Kumada, D. Terasawa, Y.Shimoda, H. Azuhata, A. Sawada,
      Z. F. Ezawa, K. Muraki, T. Saku and Y. Hirayama, Phys. Rev. 
      Lett. {\bf89}, 116802 (2002).
\item K.N. Shrivastava, Introduction to quantum Hall effect,\\ 
      Nova Science Pub. Inc., N. Y. (2002).
\item K.N. Shrivastava, cond-mat/0201232.
\item K.N. Shrivastava, cond-mat/0207391.
\item R. L. Willett et al, Phys. Rev. Lett. {\bf83}, 2624 (1999).
\end{enumerate}
\vskip0.1cm
Note: Ref.4 is available from:\\
 Nova Science Publishers, Inc.,\\
400 Oser Avenue, Suite 1600,\\
 Hauppauge, N. Y.. 11788-3619,\\
Tel.(631)-231-7269, Fax: (631)-231-8175,\\
 ISBN 1-59033-419-1 US$\$69$.\\
E-mail: novascience@Earthlink.net

\vskip0.5cm

Fig.1: Jain's model of composite fermions is sketched showing 
flux quanta attached to one electron. It turns out that after 
attaching flux quanta, the resulting quasiparticle called CF will
 be too big to have the same density as that of electrons. This 
model is found to be internally inconsistent. Some experimentalists claim to have found the CF but such claims are ``not true".
\end{document}